# XPS Supported INS and DRIFT Spectroscopy of *sp²* Amorphous Carbons.

E. F. Sheka[1*], I. Natkaniec[2], E. U. Ipatova[3], Ye. A. Golubev[4], E. N. Kabachkov[5,6], V. A. Popova[1]

[1] Peoples'Friendship University (RUDN University) of Russia, Miklukho-Maklay, 6, 117198 Moscow, Russia
[2] Faculty of Physics, Adam Mickiewicz University, Umultowska. 85, 61-614 Poznań, Poland
[3] Institute of Chemistry, Komi Science Center, Ural Branch of RAS, Pervomayskaya, 48, 167982, Syktyvkar, Russia
[4] Yushkin's Institute of Geology, Komi Science Center, Ural Branch of RAS, Pervomayskaya, 54, 167982, Syktyvkar, Russia
[5] Institute of Problems of Chemical Physics RAS, Academician Semenov, 1, 142432 Chernogolovka, Russia
[6] Scientific Center in Chernogolovka RAS, Lesnaya, 9, 142432 Chernogolovka, Russia

*Corresponding author: E.F. Sheka

E-mail: sheka@icp.ac.ru

**Abstract.** We carried out a joint analysis of the INS, DRIFT, and XPS spectra of a set of *sp²* amorphous carbons of the highest carbonization rank representing natural substances (shungite carbon, anthraxolite, and anthracite), technical graphenes (laboratory reduced graphene oxides), and industrial products (carbon blacks). It was determined, that the DRIFT spectra of the studied substances consist of two components determined by hydrogen and oxygen compositions in the circumference of graphene molecules, which represent the basic structural units of amorphic compounds. Methine groups typify the hydrogen component of natural amorphics while hydroxymethyls and methyls do the same job for the studied technical graphenes and hydroxyfurans for carbon blacks. A particular specificity of the methine-based hydrogen compositions to enhance electrooptic characteristics of the DRIFT spectrum of carbon atoms has been established. A comparable analysis of DRIFT and XPS spectra has allowed a reliable personification of the oxygen functional groups compositions of the studied amorphics resulting in a set of dependable molecular models of their basic structural units.

## 1. Introduction

Vibrational spectroscopy of black powders of amorphous carbon (AC) is not one of the widely used analytical research methods. Thus, studies on IR absorption each time encountered broadband spectra with a weakly pronounced structure, which changes upon transition from one sample to another [1-14]. The Raman scattering spectra, on the contrary, depressed with their similarity (see but a few [15, 16]), preventing a detailed study of the structure and chemical composition of the studied amorphics. Moreover, these studies were carried out mainly on occasionally selected amorphous species and each time under different conditions, due to which the obtained data acquired an individual character and were difficult to give a joint analysis. This state of vibrational spectroscopy of ACs continued for many years until the onset of a new period in the life of these amorphics — the new graphene history of old amorphous carbon. The beginning of this period can conditionally be marked by the appearance of the first results on neutron scattering by schungite carbon [17]. For the first time, both the structural parameters of the substance and spectrum of its vibrations showed that shungite carbon had a nanoscale structure, the main element of which is a flat framed graphene molecule (FGM) of the first nanometers in size, and the vibrational spectrum of the substance is the spectrum of such molecules. In fact, this study showed that shungite is a representative of a large class of molecular amorphics, electronic and vibrational spectroscopy of which are well developed [18, 19].

The next step, confirming the ACs molecular structure, was the extension of the study of neutron scattering over another type of amorphous carbon - technical graphene [20] of different origin [21, 22]. Technical graphene is a laboratory product of the reduction of graphene oxide [23]. The study showed a similarity between shungite carbon and two technical graphenes, obtained first because of chemical [24] and then thermal-shock stimulated reduction [25], thus establishing a direct relationship between FGMs of schungite carbon and reduced graphene oxide (rGO). The studies confirmed the molecular structure of the studied amorphous carbons and revealed the submicron size of the technical graphene FGMs.

The final stage of the study of structure and vibrational spectra of amorphous carbon concerns neutron scattering by a set of amorphous carbons, including anthraxolite as a representative of natural substances and products of industrial multi-tonnage production - two carbon blacks [26]. It was found that the structures of anthraxolite and one of the carbon blacks are similar to each other and to the structure of schungite carbon and the corresponding FGMs are nanometer in size. The structure of the second carbon black is similar to the structure of technical graphenes and its FGMs have a submicron size. Thus, a set of amorphous carbons with a well-defined structure of their FGMs has been available in the hands of researchers.

Natural under these conditions is the question of the FGMs chemical composition. Certain information on the amount of hydrogen and the type of bonding of its atoms to the molecules carbon cores was obtained using inelastic neutron scattering spectra. Numerous studies of industrial ACs [27-36] and technical graphenes [23] indicate a significant contribution to their composition of oxygen and other heteroatoms. However, the composition of oxygen, as well as nitrogen, sulfur and so for, containing groups remained unclear. The first attempt to clarify this issue was a comparative test of the chemical composition of a particular set of AC representatives under identical conditions [37]. The set of amorphics, already studied by neutron scattering, was subjected to analytical testing using a set of modern analytical tools, including XPS. The result of this comparative study was the first atomic models of FGMs designed in accordance with the available structural and chemical data. Now there is the turn of these models confirmation, which is the main goal of current study. Relying on the success of the multi-technique comparative study of a set of ACs performed previously, we again turn to the amorphics set that combines representatives of natural substances, technical graphenes and industrial products, structurally and chemically tested. Three kinds of vibrational spectroscopy were chosen that cover inelastic neutron scattering (INS), infrared absorption through the infrared diffuse reflection known as DRIFT, and Raman scattering. Applying to molecules, first two methods are usually used to clarify functional groups, while Raman scattering is mainly concentrated on structurally symmetric nuances. The two approaches are complex in their application, based on different concepts and working with different sets of basic data, due to which it is better to consider them separately. The results of the first approach are considered in the current paper, while results of the second one are published elsewhere [38]. In both cases, a comparative analysis of the obtained spectra, accompanied by a simultaneous analysis of the XPS spectra, made it possible to identify a set of clear molecular patterns reflecting the chemical composition and structure of the FGMs under study.

The paper is composed in the following way. The description of amorphous carbon sampling and brief account of the techniques used is given in Section 2. Section 3 is devoted to the review of INS spectra previously obtained, a new interpretation of which is suggested in the current paper. Particularities of DRIFT spectroscopy, the obtained DRIFT spectra of the set of selected amorphics, and the group frequency (GF) basis of the obtained spectra interpretation distinguishing hydrogen and oxygen components of the spectra are presented in Section 4. Hydrogen component of the DRIFT spectra is considered in Section 5 while XPS supported oxygen component is discussed in Section 6. Based on the spectra analysis, suggested molecular models of framed graphene molecules related to the studied amorphics are presented in Section 7. Section 8 accumulates core entities discussed in the paper.

## 2. Amorphous carbon sampling

Three samples of natural AC are presented by shungite carbon (ShC) (Shun'ga deposit, Karelia, Russia), anthraxolite (AnthX) (Novaya Zemlya deposit, Russia), and two anthracites (AnthC) (Donetsk coal basin, Rostov region, Russia and La Mûre deposit in France (INS only)). Four other samples are synthetic amorphics presented by two technical graphenes produced in the course of oxygenating-reducing reaction (Ak-rGO) [39] and by thermal-explosion process (TE-rGO) [40] as well as by two carbon blacks 699632 (CB632) and 699624 (CB624) produced by Merk-Sigma-Aldrich [41]. Necessary information concerning their structural and chemical content is listed in Tables 1 and 2. The data obtained earlier are supplemented in this work by the results of the X-ray diffraction for anthracite and XPS data for anthracite and two technical graphenes. A detailed description of the instruments used and the experimental data processing technique are presented in [37]. Four species, beside technical graphenes and CB624, are nanostructured and mesoporous. A general view on their structures is presented in Fig. 1. As seen in the figure, natural amorphics and CB632 represent complexes of micrometer aggregates with a lot of free space between them exhibiting the origin of the species porous structure. A close similarity of the structure appearance of all samples well support a common structural architecture of the species. Based on multilevel fractal composition of ShC [42, 43], the latter rests on the FGMs discussed earlier that represent basic structure units (BSUs) as structure elements of the first level. BSUs are composed in stacks at the second level with grouping the stacks in globules at the third level and completing the agglomeration of globules at the fourth level. The first two levels are directly supported with detailed HRTEM studies [16, 20, 44] as well as with neutron and X-ray diffraction discussed earlier. The last two levels correlates perfectly with multidimensional porous structure of the species [41, 45-47]. Structural parameters related to BSU themselves ($L_a$) and stacks of them ($L_c$) are given in Table 1. Both values are of nanometer scale thus providing nanostructuring of the samples. Important to note that the structure of CB632 is well similar to that of natural ACs [37]. In contrast, both technical graphenes and CB624 look quite differently due to its nanostructuring only in *c*-direction while the relevant BSUs are of submicron size in the lateral dimension.

**Table 1.** Structural parameters of amorphous carbons [1]

| Samples | d (Å) | $L_{CSR}^c$, nm | Number of BSU layers | $L_{CSR}^a$, nm | Ref |
|---|---|---|---|---|---|
| Graphite | 3.35 | >20[2] | ~100 | >20 | [37] |
| ShC | 3.47(n); 3.48(X) | 2,5(n); 2.0(X) | 7(n); 5-6(X) | 2.1(X) | [37] |
| AnthX | 3.47(n); 3.47(X) | 2.5(n); 1.9(X) | 7(n); 5-6(X) | 1.6(X) | [37] |
| AnthC (Donetsk) | 3.50(X) | 2.2(X) | 5-6(X) | 2.1(X) | This work |
| CB632 | 3.57(n); 3.58(X) | 2.2(n); 1.6(X) | 6(n); 4-5(X) | 1.4(X) | [37] |
| CB624 | 3.40(n); 3.45(X) | 7.8(n); 4.1(X) | 23(n); 12(X) | 2.5(X) | [37] |
| Ak-rGO | 3.50(n) | 2.4 | 7(n) | >20 | [21] |
| TE-rGO | 3.36(n) | 2.9 | 8(n) | >20 | [21] |

[1] Notations (n) and (X) indicate data obtained by neutron and X-ray diffraction, respectively.

[2] The definition ">20 nm" marks the low limit of the dimension pointing that it is bigger than the CSL of crystalline graphite equal to ~20 nm along both $a$ and c directions. Actual dimensions are of micrometer range.

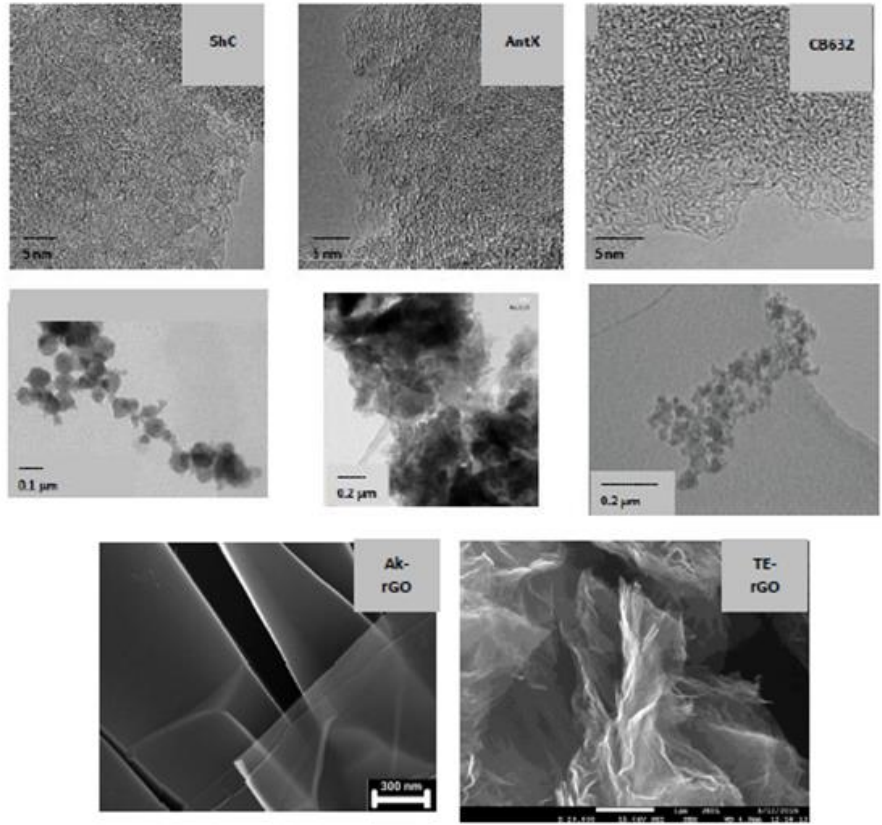

**Figure 1**. HRTEM (top), TEM (middle) and SEM (bottom) images of shungite carbon, anthraxolite, carbon black 632, technical graphene Ak-rGO and TE-rGO. Adabted from Refs. [20], [16], [41], [21], and [22], respectively.

**Table 2**. Chemical content of amorphous carbons

| Samples | Elemental analysis, wt% | | | | | Ref. | XPS analysis, at% | | | Ref. |
|---|---|---|---|---|---|---|---|---|---|---|
| | C | H | N | O | S | | C | O | Minor impurities | |
| ShC | 94.44 | 0.63 | 0.88 | 4.28 | 1.11 | [37] | 92.05 | 6.73 | **S** - 0.92; **Si** – 0.20; **N**-0.10 | [37] |
| AnthX | 94.01 | 1.11 | 0.86 | 2.66 | 1.36 | [37] | 92.83 | 6.00 | **S** - 0.85; **Si** – 0.25; **N**-0.07 | [37] |
| AnthC | 90.53 | 1.43 | 0.74 | 6.44 | 0.89 | this work | 92.94 | 6.61 | **Cl** - 0.11 - **S**: 0.34 | this work |
| TE-rGO | 84.51 | 1.0 | 0.01 | 13.5 | 1.0 | this work | 86.77 | 10.91 | **F** - 077; **S** - 0.86; **Si** -0.70 | this work |
| Aк-rGO | 89.67 | 0.96 | 0.01 | 8.98 | 0.39 | this work | 94.57 | 5.28 | **S** - 0.16 | this work |
| CB624 | 99.67 | 0.18 | 0 | 0.15 | - | [37] | 95.01 | 4.52 | **Si** – 0.46 | [37] |
| CB632 | 97.94 | 0.32 | 0.04 | 1.66 | 0.68 | [37] | 93.32 | 6.02 | **Si** – 0.66 | [37] |

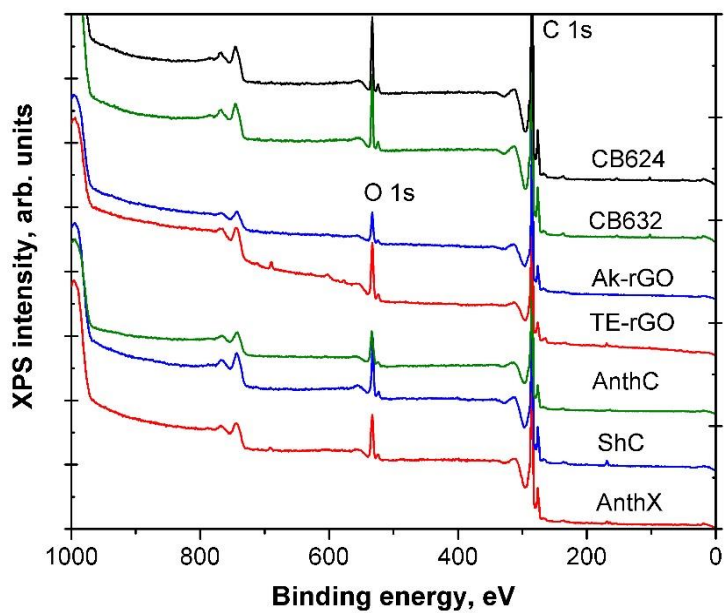

**Figure 2**. XPS survey spectra of *as prepared* amorphous carbons at room temperature.

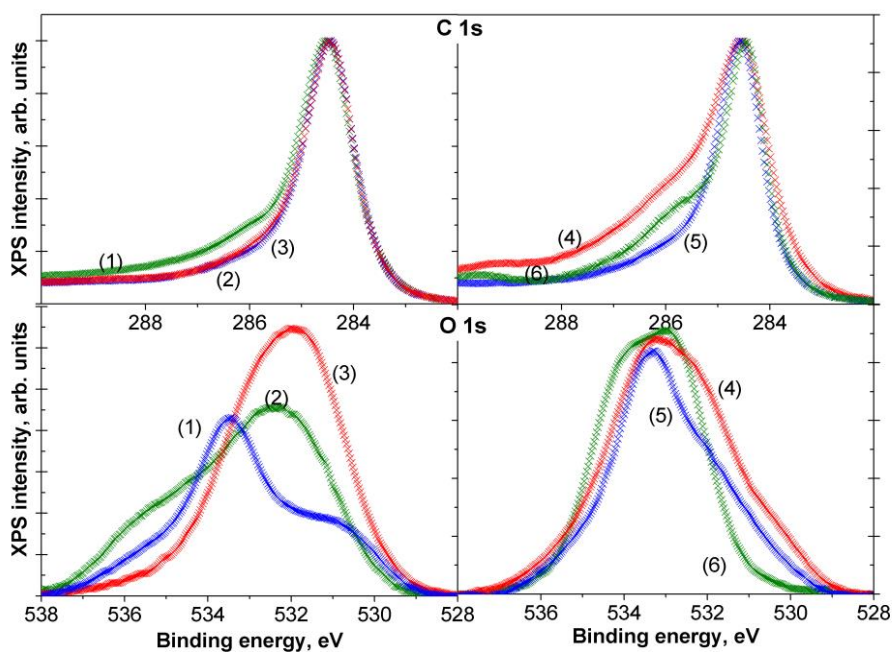

**Figure 3.** C 1s (top) and O 1s (bottom) XPS spectra of AnthC (1); ShC (2); AnthX (3); TE-rGO (4), Ak-rGO (5) and CB632 (6) at room temperature.

Data listed in Table 2 are related to the ACs BSUs although averaged over samples. The difference in the data obtained by elemental analysis CHNS and XPS is connected with spatial anisotropy of the sample BSUs discussed in [37] due to which the former data are related to the whole molecules while the latter are mainly related to the circumferences. Shown in Fig. 2, XPS survey spectra open comparative study of the AC samples. As seen in the figure, the spectra are common for all the samples demonstrating that carbon C 1s and oxygen O 1s spectra are the main contributors in all the cases while differing by atom content value. The contribution of other heteroatoms is quite small (see Table 2). The detailed difference of the spectra is much better vivid in Fig. 3. As seen, both C 1s and O 1s spectra differ quite markedly thus pointing to different compositional structure of BSUs of the studied ACs and making the issue the main target of the further discussion.

## 3. INS spectra and hydrogen content of amorphous carbons

INS spectroscopy is known as an efficient H-tool allowing establishing both the presence of hydrogen atoms in the studied samples and their chemical bonding [48]. Evidently, the 'not-true-carbon' nature of the studied amorphics, particularly their hydrogen-enrichment followed from Table 2, is highly challenging to apply for INS study of ACs. INS spectra of ShC, AntX, Ak-rGO, TE-rGO, CB624, and CB632 were obtained at the high flux pulsed IBR-2 reactor of the Frank Laboratory of Neutron Physics of JINR using NERA spectrometer [17, 21, 22, 24-26]. The spectrum of anthracite (La Mûre deposit), quantitative characteristics of which is similar to those of AnthC (Donetsk) notified in Table 2) was obtained on the TFXA spectrometer at the ISIS pulsed-neutron source, Rutherford Appleton Laboratory, Chilton, UK [49]. All the spectra were converted from counts per channel to generalized density of vibrational states (GVDOS) $G(\omega)$ per energy transfer [48] by the relevant standard programs.

Figure 4 presents a collection of GVDOS $G(\omega)$ spectra (scattering function $S(Q, \omega)$ [48] in the case of anthracite [49]) divided into two groups related to natural and synthetic amorphics, respectively. The spectra are provided with both the direct scattering from hydrogen atoms, chemically bound with the carbon cores in the BSU framing areas, and enhanced scattering from the carbon atoms of the molecule core due to 'riding effect'. The latter is caused by the contribution of hydrogen atoms to the eigenvectors of vibrations related to carbon atoms through over the molecules (see detailed discussion of both effects in [21]). The collection of NERA spectra is not full with respect to the sample list presented in Table 2 since the intensity of the CB624 spectrum under the maintained conditions is at the statistical nil level (see details in [26]). As for the other spectra, their intensity well correlates with the hydrogen content listed in the table. As seen in the figure, the two-group division of the spectra is not just formal to distinguish the different origin of the studied amorphics but, actually, presents two groups of spectra that are drastically different. Thus, all the spectra in Fig. 4a are much in common regardless of their obtaining at different facilities. Characteristic spectral features located in the region of 960 cm$^{-1}$, 800 cm$^{-1}$, and 600 cm$^{-1}$ are clearly vivid in all the spectra thus indicating a similar involvement of hydrogen atoms in the scattering. A profound analysis of the involvement showed that the first two features are attributed to the in- and out-of-plane bending vibrations (*ip* and *op* bendings, respectively) of C-H bonds of methine groups [17, 25, 26, 49], while the third one represents *op* bendings of carbon atoms of the benzenoid units of the graphene-like cores [50] enhanced by 'riding effect'. The spectra below 500 cm$^{-1}$ are related to *ip* bendings of carbon atoms supplemented with low-frequency acoustics, the appearance of which is due to the presence of hydrogen atoms via "riding effect" as well. Therefore, the INS study of natural ACs reveals that in addition to the similarity in molecular spatial structure of their BSUs, hydrogen atoms in their framing areas are predominantly chemically bound forming sets of methine $sp^2$C-H bonds.

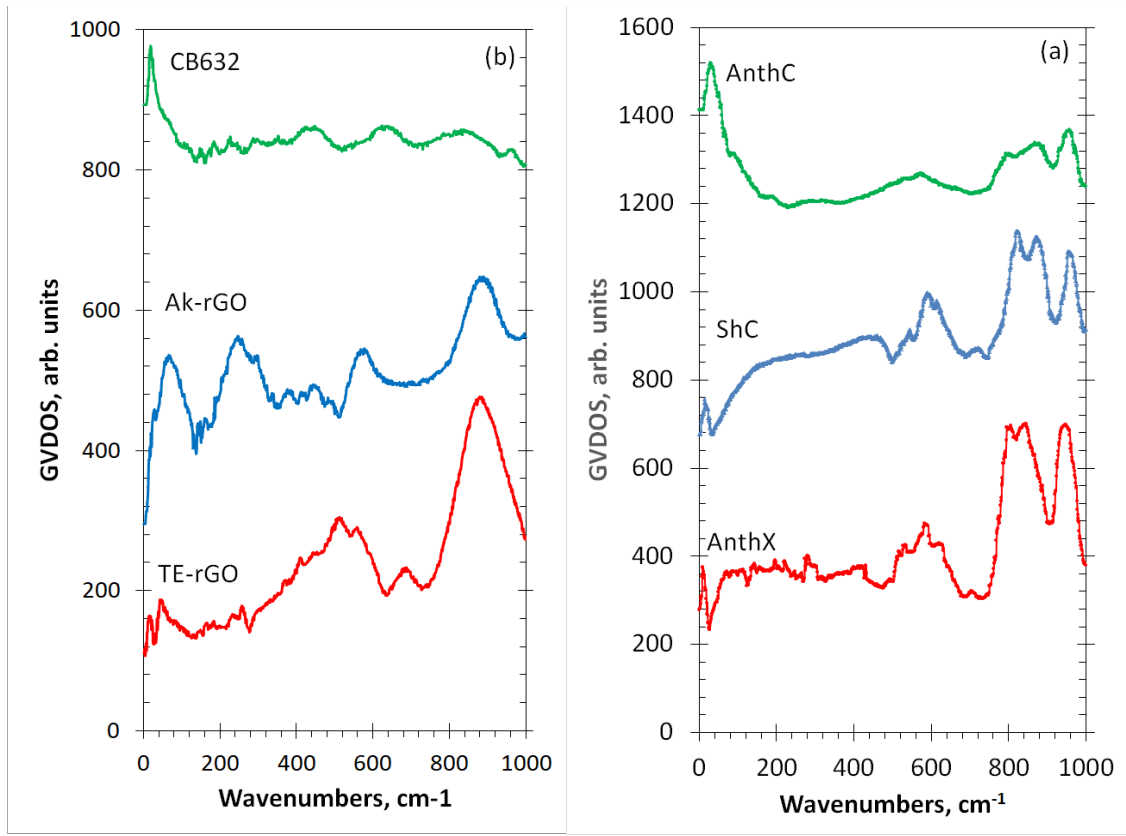

**Figure 4.** Experimental GVDOS $G(\omega)$ spectra of *sp²* amorphous carbons at T=20K: (a) - AnthX [26]; ShC (600) [24]; AnthC (1000) [49]. (b)–TE-rGO [24]; Ak-rGO (200) [24]; CB632 (800) [26]. Digits in brackets indicate upshift along vertical axis.

In contrast to natural amorphics, their synthetic analogues are characterized by quite different $G(\omega)$ spectra as seen in Fig. 4b and should be interpreted separately. At first, there is a marked difference between technical graphenes and carbon black: if the former are sufficiently hydrogen enriched, the latter are hydrogen poor. Referring to the calculated H-standard INS spectra [26] it is possible to attribute the main features of the Ak-rGO spectrum at 100 cm$^{-1}$, 220 cm$^{-1}$, 600 cm$^{-1}$ and 860 cm$^{-1}$ to the presence of CH$_3$ groups in the framing area of the sample. Actually, a highly broad peak at 860 cm$^{-1}$ evidently substitutes three peaks of the spectra of natural amorphics in the region of 800-960 cm$^{-1}$. Such a behavior is usually typical for the interrelation of *sp²*C-H, on the one hand, and either *sp²*C-CH$_2$ or *sp²*C-CH$_3$ groups, on the other, in the vibrational spectra of organic molecules [51]. The dominant role of CH$_3$ groups is supported by the presence of peaks at 100 cm$^{-1}$ and 220 cm$^{-1}$ that are undoubtedly attributed to CH$_3$ rotations and bendings. The remaining peak at 600 cm$^{-1}$ is evidently attributed to the bendings of the core carbon atoms sensitized with hydrogen atoms by riding effect. This interpretation of the vibrational spectrum of the Ak-rGO species is consistent with the character of chemical reaction laying the foundation of the species production [39]. Actually, the reduction of graphene oxide is accompanied with the propyl alcohol decomposition leading to the presence of methyl radicals in the chemical surrounding. Naturally, the termination of dangling bonds of graphene fragments on the way of the relevant BSUs formation is readily performed by the addition of CH$_3$ groups.

When going to the $G(\omega)$ spectrum of TE-rGO, one faces the fact that characteristic peak at 860 cm$^{-1}$ not only remains, but even becomes relatively stronger while peaks at 100 cm$^{-1}$ and 220 cm$^{-1}$ disappear. Such a behavior is characteristic for organic molecules when *sp²*C-CH$_3$ groups are substituted with *sp²*C-CH$_2$ ones [ 51]]. Simultaneously, *sp²*C-CH$_3$ bendings at 220 cm$^{-1}$ shift up to frequencies of *sp²*C-CH₂ groups and provide the appearance of new peak at 500 cm⁻¹ that, in their turn, stimulate upshifting the bendings of the core carbon atoms to 640 cm⁻¹. The presence of *sp²*C-CH₂ groups in the framing area of TE-rGO product is highly expected [52]. Actually, one of the intermediate procedures on the way to thermally reduce the pristine graphene oxide (GO) consists in placing GO as a paste on either polyethylene or polypropylene films. After heating at moderate temperature, GO film is taken off from polymer but some residual amount of the latter may be put together with GO in a furnace. Thermal shock at T=750⁰ C causes unavoidable destruction of polymer thus supplying the emerging rGO with methylene radicals to terminate active dangling bonds of the BSUs edges. Therefore, the $G(\omega)$ spectra of both technical graphenes clearly evidence the role of chemical surrounding for the chemical production of the framing area of BSUs of the relevant products. Evidently, the feature is responsible for a large variety of chemical composition of technical graphene produced by different ways [23]. This chemically assisted explanation of the BSU framing has been directly proven for technical graphene in the current study for the first time.

The presence of *sp²*C-CH₃ and *sp²*C-CH₂ groups in the BSU framing areas of Ak-rGO and TE-rGO species evidently provide high intensity of scattering. Oppositely, the INS spectrum and consequently $G(\omega)$ spectrum of CB632 in Fig. 4b, is very weak while that one of CB624 cannot be distinguished at all. The feature is evidently connected with high-temperature pyrolytic conditions of the amorphic production, which does not maintain the existence of hydrogen-containing radicals in the chemical surrounding. As mentioned before, the hydrogen content in CB632 is at the limit of the technique sensitivity due to which only a sharp feature below 100 cm⁻¹ marks the presence of hydrogen atoms in the species BSU circumference. As will be shown below, it might be attributed to hydroxypyran compositions in the areas.

Summarizing the comparative analysis of the INS spectra of the studied ACs, we can draw the following conclusions.

i. Studied natural amorphics contain hydrogen at the level of 1±4 wt%. The hydrogen atoms are located in the framing area of the amorphics BSUs and terminate the molecules edge carbon atoms by the formation of *sp²*C-H bonds. Such a universality concerning so different massive natural products convincingly evidences the existence of common fundamental laws governing the origin of graphene lamellae in the Nature as well as the presence of water as the only source of hydrogen in the species chemical surrounding. It must be borne in mind that here we are dealing with a kinetically completed process over a billion to a million years. As occurred, methine *sp²*C-H groups are the most stable.

ii. Due to complex hydrogen-rich chemical environment which accompanies the species production, all technical graphenes are (and must be) hydrogen enriched at the level up to 1-2 wt%. The type of hydrogen containing units, which decorate the BSU framing areas, depends on concomitant reagents participating in the reduction process. In the current case, these are CH₃ and CH₂ radicals provided by the presence of propyl acohol and alkene polymers when producing Ak-rGO and TE-rGO, respectively.

iii. The hydrogen content of synthetic carbon blacks is too low due to which its reliable attribution to particular chemical units required using additional techniques.

## 4. DRIFT spectroscopy: Spectra and group frequency basis

*"The qualitative aspects of infrared spectroscopy are one of the most powerful attributes of this diverse and versatile analytical technique. Over the years, much has been published in*

*terms of the fundamental absorption frequencies (also known as **group frequencies**) which are the key to unlocking the structure–spectral relationships of the associated molecular vibrations. Applying this knowledge at the practical routine level tends to be a mixture of art and science"*- John Coats [53]. Fully sharing this viewpoint of a professional expert, we follow the approach suggested by the author based on the opinion that the main goal of IR spectra interpretation is not the assigning as many bands as possible using group frequency (GF) tables. More important is to build up a self-consistent picture using information from both key spectral bands and any known history, including the sample preparation and presentation technique. The approach particularly suits the case of highly complex DRIFT vibrational spectra of amorphous carbons. We will begin this analysis primarily with an understanding of what image of the experimental DRIFT spectrum we are dealing with.

As well known (see a comprehend review [54] and references therein), Kubelka-Munk (KM) data [55] provided with the standard DRIFT device software present the reflectance and are quite sensitive to a number of external parameters. The latter involves such values as size of powder particles, their pristine packing in the powder, on the one hand, and the powder packing in the device sample cup, on the other. Additionally, any structural and/or geometrical perturbations, which disturb the isotropic reflectance, linear relationship between KM function and absorbance, such as constancy of reflection coefficient, as well as those that stimulate the appearance of specular reflectance, are significant. Accordingly, KM spectra are highly changeable and usually serial measurements for each sample are needed to get reproducible spectral image. However, even then DRIFT spectra, based on measuring the reflectance, may drastically differ from the transmittance spectra and do not present exact absorbance of the samples [54, 56] similarly to that as the X-ray photo of a human body is not identical to the body portrait obtained in visible light. In view of this, the DRIFT spectra exhibit the vibrational spectra of the samples partially, concerning mainly frequencies only while their contributions to the total spectrum intensity are not governed by the relevant dipole moments and their derivatives, as usually accepted [57]. In spite of the shortcomings, DRIFT spectroscopy becomes a valuable tool when a set of samples are studied under the same conditions just forming the ground for a reliable comparative study. In the current study, DRIFT spectra of seven ACs were collected at room temperature using Fourier transform infrared device IRPrestige -21, Shimadzu equipped with diffuse reflectance unit DRS-8000 in the regime of 4 $cm^{-1}$ resolution, 40 scans and digital averaging over 15 points. The reference reflectance was registered from the provided mirror. Samples were carefully grinded before each measurement and put in the device cup in pure form.

Figure 5 presents a complete collection of the DRIFT spectra obtained. It should be noted small KM values in all the cases evidencing high absorbance of the samples. When presenting, the spectra were divided into three groups with relation to the amorphics origin. As seen in the figure, this division turned out quite appropriate clearly evidencing a similarity between the spectra within each group, including both the KM data scale (intensity) and spectral features, as well as allowing at the same time to see both common and different features of the spectra of different groups. Before to start spectra interpretation, it should be borne in mind that, firstly, the predominant carbon content is common to all samples, constituting from 84.5 to 99.7 wt% according to Table 2. Secondly, BSUs size of the samples of the first group and CB632 is ~1.5-2.0 nm. For samples of the second group and CB624, the value is much bigger (see Table 1). Consequently, the number of carbon atoms is of 150-200 in the first case and an order of magnitude bigger in the second while the contribution of framing atoms is of first at% in all the cases. Accordingly, it was to expect that the spectra of the studied samples should contain a significant contribution from the vibrations of the carbon skeleton, which is significantly similar for all the samples, thereby causing at least a partial similarity of the spectra. As seen in Fig. 5, a clearly visible similarity is, actually, characteristic for natural ACs while no trace of anything like that is observed when comparing the spectra of natural species with those of technical graphenes and carbon blacks. This feature alongside with drastic lowering of the spectra intensity in the latter two cases convincingly evidences that not carbon atoms of the BSUs cores but those of the framing areas play the governing role in the spectra patterns of these ACs. Thus, a general comparative analysis of the obtained spectra convincingly indicates a deep connection between the vibrational dynamics of the studied substances and the chemical history of their preparation. It is this important conclusion that will be a leading point of the further interpretation of the obtained spectra. At the same time, the interpretation described below heavily relies on the available GFs data bank obtained for aromatic molecules and establishing a relationship between the structural elements and particular vibrational modes. However, in the case of the studied ACs with several hundreds of the modes, these relations do not apply to individual modes, but concern large groups of the latter thus marking frequency regions only.

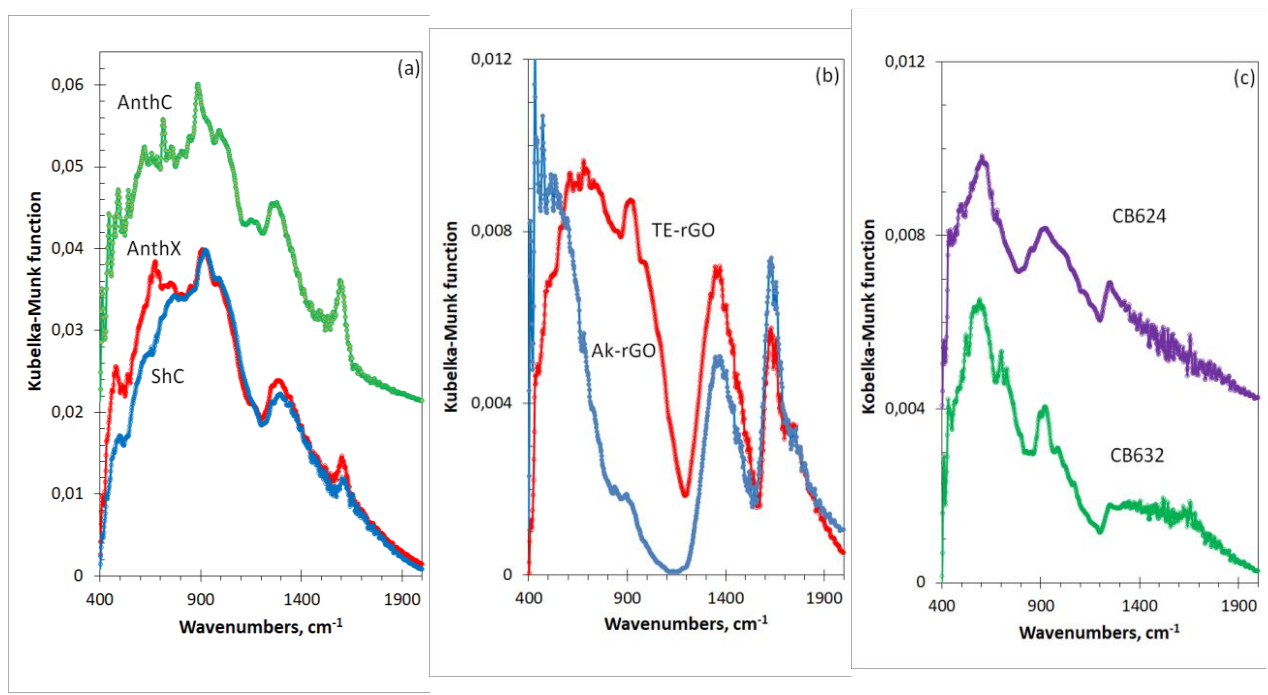

**Figure 5.** DRIFT spectra of amorphous carbons at room temperature: natural ACs (a), technical graphenes (b), and carbon blacks (c). Within each group, spectra are normalized per maximum. The vertical shift of spectra is 0.02 (a), 0 (b), and 0.004 (c).

Vibrational spectroscopy of aromatic molecules convincingly indicates that GFs of benzene and its derivatives can be considered as a standard basis for interpreting the spectra of molecules with a complex structure [50, 51, 58]. We will take advantage of this circumstance in the present case, especially since we are not talking about individual modes, but rather wide vibrational intervals. As seen in Fig. 5, DRIFT spectra of all studied ACs are broadband and cover wide region from 400 to 2000 $cm^{-1}$. Aiming at the spectra interpretation in terms of GFs and following the data presented in Table 2, which show a complicated chemical composition of the studied BSUs, BSU molecules DRIFT spectra should be presented in terms of (C, C), (C, H), (C, O), (O, H), and (C, H, O) vibrations. Therefore, we are facing the case when many ambiguities and coincidences related to the relevant GFs are expected. The only way to cope with this difficulty is to sequentially analyze the DRIFT spectra in the light of individual (C, X) atomic groups (X=C, H, O), the completion of which should be carried out taking into account other independent data obtained by other methods or concerning the history of substances. In this work, the INS and XPS were chosen as particular arbiters. Table 3 accumulates the results obtained and simultaneously illustrates the analysis procedure. The first column of the table presents the characteristic regions of the DRIFT spectra. Next columns provide a possible interpretation of these regions in terms of GFs of the corresponding (C, X) atomic groups.

**Table 3.** Standard group frequencies of aromatic molecules required for the fractional analysis of vibrational spectra of amorphous carbons

| DRIFT regions, cm$^{-1}$ | Group frequencies[1] | | | | | |
| --- | --- | --- | --- | --- | --- | --- |
| | (C, C)[2] | (C, H$_1$)[2] | (C, CH$_2$)[3] | (C, CH$_3$)[4] | (C, O$_1$) and (C, OH)[5] | (C, O-C) |
| 400-700 | 404 δ *op* C-C-C<br>606 δ *ip* C-C-C | - | 711 ρ CH$_2$ | 210 *r* CH$_3$<br>344 δ CH$_3$ | 458 τ COH | 605 δ C-O |
| 700-1200 | 707 C-C-C puckering,<br>993 ring breathing,<br>1010 δ C-C-C trigonal | 673 δ *op* in phase<br>846 δ *op*, C$_6$ libration<br>967 δ *op*<br>990 δ *op*, trigonal<br>1037 δ *ip*<br>1146 δ *ip*, trigonal<br>1178 δ *ip* | 948 ρ CH$_2$ | 900 ν C-CH$_3$<br>1041 ρ CH$_3$ | 960 ν C-O(H)<br>1158 δ C-O(H)<br>1284 ν C-O(H) | 970 ν C-O<br>1260 ν C-O |
| 1200-1600 | 1309 ν C-C Kekule,<br>1482 ν C-C<br>1599 ν C-C | 1350 δ *ip* in phase | 1409 δ internal CH$_2$ | 1333 δ CH$_3$<br>1486 δ internal CH$_3$ | 1511 ν C=O | - |
| 1600-1900 | - | - | | | | |
| 2800-3200 | - | 3056 ν C-H<br>3057 ν trigonal C-H<br>3064 ν C-H<br>3073 ν in phase C-H | 3114 ν CH$_2$ | 2950 ν CH$_3$ | 3400 ν OH | - |

[1] Greek symbols ν, δ, ρ, *r*, τ mark stretching, bending, rocking, rotational, and torsion modes, respectively.
[2] GFs notifications of fundamental vibrations of benzene molecule [50].
[3] GFs notifications of fundamental vibrations of benzyl radical [59, 60--]. Hereinafter, GFs, additional to the benzene pool of vibrations, will be shown only.
[4] GFs notifications of fundamental vibrations of toluene [61, 68].
[5] GFs notifications of fundamental vibrations of *p*-benzosemiquinone [63].

### 4.1. GF marking of (C, C) vibrations

Obviously, the consideration of (C, C) vibrations, which present the main massif of the ACs vibrational modes, opens the GF fractional analysis. As virtually shown [64], (C, C) vibrations of honeycomb nanosize composition of carbon atoms densely cover region from a few tens to ~1600 cm$^{-1}$. Figure 6 just illustrates the said and presents 564 vibrational modes of a non-terminated (9, 9)NGr molecule (see insert) of ~2 nm in size, consisting of 190 carbon atoms. These modes fill the frequency space under the observed broad band. Evidently, there is no possibility to describe the mode set with a limited number of the relevant GFs due to which it is reasonable to apply for a qualitative classification to standard vibrational modes of benzene, exhaustedly studied experimentally and computationally [50, 58]. (C, C) vibration of the molecule are shown in the figure by red spikes. As occurred, such a classification is quite reliable due to evident spreading of each mode of benzene into a well separated group of modes related to graphene molecule. The (C, C) frequency pool of benzene is usually divided into three main parts, namely,

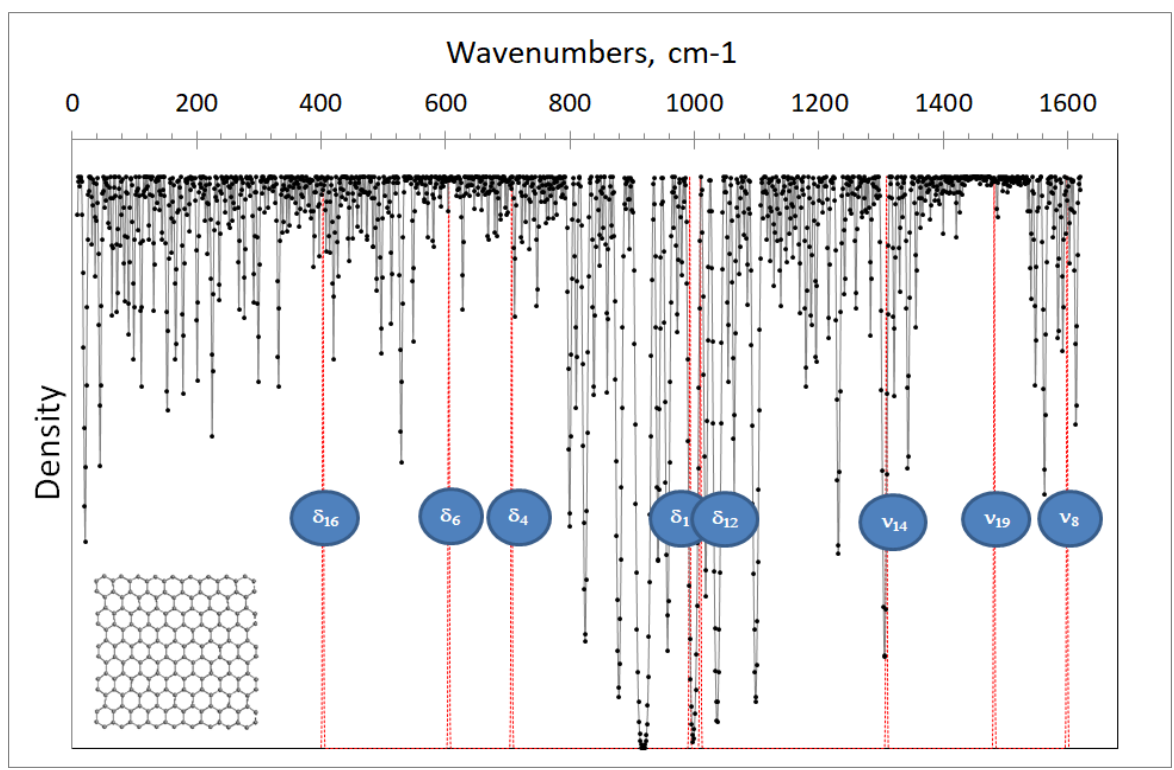

**Figure 6.** Calculated density of vibrational modes of (9, 9)NGr (black) and benzene (red) molecules adapted from [50]. Calculations were performed in the framework of the UHF AM1 semi empirical approximation (see details in [60]). δ-Functions of the same total intensity, presenting vibrational modes, were convoluted with Lorentzians of 10 and 2 cm$^{-1}$ FWHM for graphene and benzene molecules, respectively. Circle labelling presents standard attribution and numbering of benzene molecule modes [50].

400 - 700 cm$^{-1}$ (I), 700 - 1200 cm$^{-1}$ (II), and 1200 - 1600 cm$^{-1}$ (III). As seen in the Fig. 6, Part I, originated from *op* and *ip* C-C-C bendings, is transformed into a broad spectral region from ~10 to 700 cm$^{-1}$ in the benzenoid-condensed graphene molecule. Part II, stemmed from the C-C-C trigonal bending, C-C-C ring breathing, and C-C-C puckering, originates two strongly dispersed groups of modes when the rings are condensed. Part III, attributed to three groups of C=C stretchings, in graphene molecule are presented by three broad sets of modes, with the most dispersed $\nu_{14}$ Kekule ones. The discussed benzene-based GFs of (C, C) vibrations of graphene molecules are summarized in Table 3.

### 4.2. (C, H) vibrations

The next step of the GF fractional analysis concerns (C, H) vibrations. According the INS study discussed in the previous section, in the current study, we are dealing with three types of the relevant chemical bonding and must consider three sets of GFs related to $sp^2$C-H, $sp^2$C-CH$_2$, and $sp^2$C-CH$_3$ bonds. The corresponding GF sets are notified in Table 3 as (C, H$_1$), (C, CH$_2$), and (C, CH$_3$), respectively. Large experience obtained in the study of the IR spectra of aromatic molecules shows that in all the cases when these functional groups are present in the molecule composition, the corresponding vibrations govern the spectrum intensity [51, 58, 65-69]. From this standpoint, it is reasonable to connect the drastic lowering of the intensity of the DRIFT spectra of carbon blacks compared to other ACs studied with their negligible hydrogenation, which is consistent with the INS study. Thus, only natural amorphics and technical graphenes are main subjects to analysis from the viewpoint of the (C, H) fractionation. A drastic difference in the shape of DRIFT spectra of the samples of both groups strongly validate their analysis from the above (C, H) triad standpoint.

*(C, H$_1$) vibrations.* The addition of hydrogen atoms into the BSU molecules circumferences is followed with the formation of methine groups with $sp^2$C-H bonds, similar to those of benzene. The corresponding eight vibrational modes are shown in Table 3. In benzene as well as in other aromatic molecules, the main manifestation of these bonds is associated with *op* and *ip sp$^2$*C-H bendings located in the 700-900 cm$^{-1}$ and 1000-1200 cm$^{-1}$, respectively [51, 68]. As for vibrational spectra of the studied samples, a similar feature can be found in the DRIFT spectra of natural amorphics.

*(C, CH$_2$ ) vibrations.* This and the next (C, H) vibration groups are related to technical graphenes. If the choice of benzene and its GFs as the standard in the previous case was invariant, then in the case of (C, CH$_2$) vibrations, we are dealing with a large number of the benzene molecule derivatives, in the structure of which there are $sp^2$C-CH$_2$ bonds. To simplify the selection, we address our understanding of the chemical reaction accompanying the preparation of technical graphene TE-rGO. As follows from the chemical procedure of the species production [52], dangling bonds on the edge atoms of the exposed graphene lamellas, obtained in the course of the reduction of GO molecules under the temperature shock, are terminated by CH$_2$ radicals. This results in the formation of benzyl-radical structures at the edges of this amorphic BSUs. Consequently, it is natural to choose a benzyl radical molecule as the source of the required GF standards. Listed in Table 3 is a set of GFs, related to vibrations attributed to $sp^2$C-CH$_2$ bonds of the molecule [59, 60]. The set is surrounded with a large massive of benzene data presented in columns (C, C) and (C, H). However, IR spectrum of the molecule is very simple and is presented with three (C, CH$_2$) bands indicated in the table which strongly dominate over low-intense extended background [59].

*(C, CH$_3$) vibrations*. As evidenced by INS study, the production of Ak-rGO technical graphene actually occurs in the chemical surrounding involving CH$_3$ radicals [39]. Naturally expected, toluene should be considered as the first candidate for the required GF standard. Vibrational spectrum of toluene was widely examined, although certain ambiguity concerning vibration modes still remains [58, 61, 62]. Nevertheless, reliable GFs related to $sp^2$C-CH$_3$ bonds can be obtained (see Table 3).

### 4.3. (C, O) vibrations

Oxygen takes an active part in the chemistry and structure of aromatic molecules, forming entire classes of the derivative molecules: phenols, lactones, quinones and hydroquinones, pyrans and hydroxypyrans, cyclic ethers, and many others. Nevertheless, describing this manifold from the GFs standpoint, one manages to limit oneself to quite small number of standards required [51, 53]. We consider three of them that might have a direct connection to amorphous carbons judging by the nature of the chemical processes that accompany their production when edge benzenoid units themselves or their hydrogenated derivatives are objects of a further oxidation.

*(C, O$_1$) and (C, OH) vibrations*. The considered GFs concerns vibrations related to oxygen atoms located outside benzenoid rings. The former type of vibrations concerns C=O carbonyl units either themselves or incorporated in more complex structures such as quinones, lactones, acid anhydride as well as carboxyls. In spite of large number of possible molecular structures, the corresponding set of GFs usually comes down to the only standard related to stretching vibrations of the C=O type. Depending in the molecules, this standard GFs cover range from 1520 to 1700 cm$^{-1}$ [51, 53]. The second GF type concerns hydroxyls attached to either pristine or modified benzenoid rings. Similarly to the previous case, in spite of a great number of different hydroxylation possibilities, the corresponding GFs set is quite standard. Presented in Table 3 is related to *o*-benzosemiquinone molecule that is one of the best cases to show both types of GFs simultaneously [63]. The data are typical for these kinds of bonds in aromatic molecules of different complexity with the frequency deviations not more than a few % [51, 53].

*(C, O-C) vibrations* mark C-O-C structures of pyran and cyclic ester originated with the incorporation of oxygen atoms into edge benzenoid rings of graphene molecules. As well known, the triplet of GFs is highly characteristic for cyclic esters, for which bands at 1260 and 970 cm$^{-1}$ present a highly characteristic doublet of the C-O stretchings while that one at 605 cm$^{-1}$ presents C-O deformations [51]. The corresponding GFs are introduced in Table 3.

Generally, to complete the composition of standard GFs related to amorphous carbons is necessary to complement Table 3 with the relevant data attributed to (C, N), (C, S) and other groups involving heteroatoms [1, 27, 30, 35]]. However, according to Table 2, the contribution of these impurities is quite negligible due to which we can limit ourselves by the considered atomic triad C-H-O and utilize the data of Table 3 to build up a self-consistent picture of the DRIFT spectra of the studied ACs. Following the concept of fractional analysis, we divide the further consideration into two parts related to the hydrogen and oxygen components of the studied DRIFT spectra. As seen in Table 3, both components cover the whole region of the spectra. Nevertheless, two-component consideration of IR spectra of aromatic molecules allow revealing a governing and additional role of the first and second components, respectively [51, 53]. This observation is fully supported in the studied DRIFT spectra of ACs. Thus, DRIFT spectra of carbon blacks in Fig. 5c are the least intense due to negligible and/or nil hydrogenation of the species shown by INS. They may exemplify the contribution of the oxygen component. On the other hand, DRIFT spectra of technical graphenes are twice stronger than the spectra of carbon blacks evidencing practically equal contributions of both components. The DRIFT spectra of natural amorphics are obviously predominantly hydrogenous. Usually, the detection of hydrogen atoms is confirmed by evidencing C-H stretchings. As seen in Table 3, the relevant GFs cover region from 3110 to 2950 cm$^{-1}$. Unfortunately, no bands are observed in the obtained DRIFT spectra. The same concerns O-H stretchings of the adsorbed water, always present in the studied amorphics (see review [25]) that are not detected in the 3200-3600 cm$^{-1}$ region as well. Apparently, in the first case, the feature is connected with a condensed polybenzenoid structure of the BSU molecules since, as known [52], the intensity of IR spectra in this region drastically lowers in polyaromatic hydrocarbons when the rings number increases. The reasons for the OH bands absence has not been so far clear. However, the ability of DRIFT technique to miss valuable bands should be mentioned.

The absence of C-H valence vibrations bands in the DRIFT spectra undoubtedly complicates the interpretation of the spectra, since, as can be seen from Table 3, in the 400–1600 cm$^{-1}$ region GFs of different functional groups are close or coincide. As a result, any information regarding the chemical history of the samples is of utmost importance. An additional help is the lack of symmetry of the experimental samples, which eliminates the consideration of the influence of symmetry selection rules on the spectra formation.

## 5. Hydrogen component of DRIFT spectra of amorphous carbons

### 5.1. Natural amorphics

As seen in Fig. 5a, DRIFT spectra of all natural amorphics in the region from 400 cm$^{-1}$ to 1600 cm$^{-1}$ look quite similar and are presented by a single broad band with moderately expressed structure. According the Table 3, three types of GFs, which are *op* and *ip* CH bendings located at ~700 cm$^{-1}$ and ~1000 cm$^{-1}$ and CH *ip* bendings in phase at 1350 cm$^{-1}$ provide the main contribution. Obviously, DRIFT spectra of the studied sample are the most intense in this region thus fully assuming that the (C, H$_1$) GFs from Table 3 manifest themselves in this area. The first two frequency regions are obviously overlapped and form the main body of the absorption bands. The same concerns broad asymmetric bands at 1300 cm$^{-1}$. Typically, all these three types of (C, H) vibrations are exhibited in IR spectra of aromatic molecules as separate bands. In contrast, the absence of distinct bands and the presence of a single convolution-like bell-shaped envelope evidences a distinguishable feature of the DRIFT spectra of the studied BSU molecules. It is difficult to propose another reason for the absence of distinct gaps between the bands related to *sp$^2$*C-H bonds, except for filling them with a continuous spectrum of (C, C) vibrations. This circumstance, apparently, is also related to the difference in the details of the DRIFT spectra shape of AnthX and AnthC in comparison with that of ShC in the region of 400-800 cm$^{-1}$. Since in this case the matter is about the region of C-C-C puckering distributed throughout the carbon skeleton of the BSU molecules, the observed difference may indicate the difference in the BSU core structures as a whole. Thus, we conclude that the observed DRIFT spectra of natural amorphics, in contrast to aromatic molecules, are formed by the joint participation of (C, C) and (C, H) vibrations. The feature is not surprised and is quite typical for aromatic molecules. Surprising is extremely high value of the carbon atoms contribution, observed on example of BSU molecules of natural amorphics for the first time.

### 5.2. Technical graphenes and carbon blacks

The transition to technical graphenes (TGs) confronts us with completely different spectra: KM signals become much lower and the DRIFT spectra are fully reconstructed. The first reaction is to associate these changes with the difference in the size of BSUs of natural ACs and those of TGs. Actually, the lateral dimensions of TG BSU molecules is bigger by an order of magnitude. However, so drastic changes are not typical for nanoparticle of AC, which was confirmed by numerous studies of IR absorption of graphene quantum dots [70, 71]. Therefore, the only remaining reason concerns the chemical configurations of the BSUs framing areas. Before starting the analysis, recall that according to the INS study, natural amorphics, on the one hand, and TE-rGO and Ak-rGO, on the other, are just *sp$^2$*C-H, *sp$^2$*C-CH$_2$, and *sp$^2$*C-CH$_3$ replicas of the vibrational spectra of the studied ACs. DRIFT spectra analysis of natural amorphics performed in the previous section confirmed the *sp$^2$*C-H status of the species suggested by INS. As for TE-rGO, its spectrum in Fig. 5b becomes more structured revealing a distinct gap at 1200 cm$^{-1}$ and the appearance of well-seen maximum at 1000 cm$^{-1}$. The spectrum sharpening as well as almost fivefold decrease of its intensity are evidently caused by disappearing a continuous component typical for natural amorphics. Leaving the explanation for this feature for later, carry out the analysis of the sufficiently structured DRIFT spectrum basing on the (C, CH$_2$) GFs set presented in Table 3.

As follows from the table, the GFs-governed IR spectra should consist of two bands of CH$_2$ rockings located at 700 and 950 cm$^{-1}$, supplemented with the band at 1400 cm$^{-1}$ of internal CH$_2$ bendings. Such a structure is typical for the IR spectrum of benzyl molecule and evidently suits the general image of the DRIFT spectrum of TE-rGO. However, the amorphic spectrum bands are much broader than in the molecule spectrum due to which the gap between two rocking bands is greatly smoothed. There might be two reasons for the broadening. The first concerns a considerable dispersion of the amorphic vibrational modes with respect to the limited modes number of benzyl molecule. The second is caused by complex character of the DRIFT spectrum, presenting a combination of hydrogen and oxygen components, due to which a GFs coincidence of simultaneous participation of (C, H) and (C, O) vibrations can be expected according to the data in Table 3. Thus, assuming it is obvious that the 400-1400 cm$^{-1}$ part of the considered DRIFT spectrum is consistent with the status of $sp^2$C-CH$_2$ replica, we leave the final interpretation of the TE-rGO BSUs spectrum to the examination of its oxygen component presented in the next section.

Moving to the Ak-rGO DRIFT spectrum in Fig.5b we see that the main changes concern the spectrum part below 1200 cm$^{-1}$. A broad band in TE-rGO spectrum discussed above contracts significantly and its maximum evidently shifts down. As follows from Table 3, this transformation concerns mainly the hydrogen component of the spectrum and is caused by transition from benzyl-like BSU framing of TE-rGO to toluene-like one in the current case. Actually, the spectrum of solid Ak-rGO is much similar to that one of toluene molecule [61, 62] due to low-frequency rotations and bendings of CH$_3$ units occupy a leading position in the DRIFT spectrum while the corresponding C-CH$_3$ rocking and stretching form a pedestal for low-frequency GFs. Additionally, two bendings at ~1400 cm$^{-1}$ form a well separated band. Therefore, the status of $sp^2$C-CH$_3$ replica is evidently supported for the hydrogen component of Ak-rGO DRIFT spectrum. Its part related to the oxygen contribution will be considered in the next section.

To complete the consideration of hydrogen components of the studied DRIFT spectra, we come back to the drastic difference in the intensity of spectra of natural ACs and TGs caused by the continuous component related to the (C, C) vibrations. We suppose that the feature represents a peculiar "riding effect" connected with a contribution of hydrogen atoms into the electrooptics of carbon vibrations. Until now, the commonly accepted consideration of electrooptics is based on the assumption of superpositional contributions of chemically different atoms to the total vibrational spectrum of polyatomic molecules [57]. This approximation, applied to aromatic molecules, allows distinguishing functional groups that are either 'free' or closely related to benzenoid rings. As empirically shown, the superpositional character of the spectra is characteristic for free functional groups to the most extent. The reason for the violation of the superposition in the opposite case is the delocalization of the electronic properties of the graphene molecule, which are uniquely sensitive to the distribution of C = C bonds. Actually, the slightest violation, such as the attachment of one hydrogen atom to one of the edge carbon atoms of the molecule, causes a significant rearrangement of the bonds [72]. The oscillatory motion of atoms in a molecule is always collective, as evidenced by 3N dimensional eigenvectors. However, if in the vast majority of cases, this collectivity is formal; in the case of graphene molecules, it constitutes the motion essence. And the special role of hydrogen atoms becomes obvious, since they are characterized by the largest displacements from the equilibrium positions. From this viewpoint, methylene and methyl of benzyl radical and toluene molecules are free since they are separated from the carbon core by additional C-C bonds thus providing a weak interaction of these groups with the benzene and/or benzenoid ring. In contrast, atomic hydrogen of methine groups takes active part in the vibrational motion of benzenoid carbon atoms.

The drastic difference in the spectra intensity and shape of carbon blacks in Fig. 5c from those in Figs. 5a and 5b is a convincing confirmation of a negligibly small contribution of hydrogen atoms in the BSU configurations of the amorphics. The feature is well consistent with the data listed in Table 2 and the INS study presented in Section 3. Possible ignorance of hydrogens significantly facilitates the spectra interpretation, concentrating the latter on the oxygen contributions only that is usually concerns mainly to GFs related to the (C, O), (C, OH), and (C, O-C) listed in Table 3. However, in order to carry out the analysis of the oxygen component most accurately, we will consider DRIFT spectra alongside with XPS ones.

**6. XPS spectra and oxygen content of amorphous carbons**

Similarly to INS, XPS spectra are highly atom-sensitive due to which they are usually considered as direct O- , N- or any other heteroatom- tools when detecting chemical content of carbonaceous matters. In our case, N- and S-components of the XPS spectra are weak due to which our interest concerns the oxygen content, which is the main contributor after carbon of ACs XPS spectra as seen in Fig. 2. All the amorphics considered in the previous section were subjected to the XPS study performed under the same conditions. Details of the experiment is

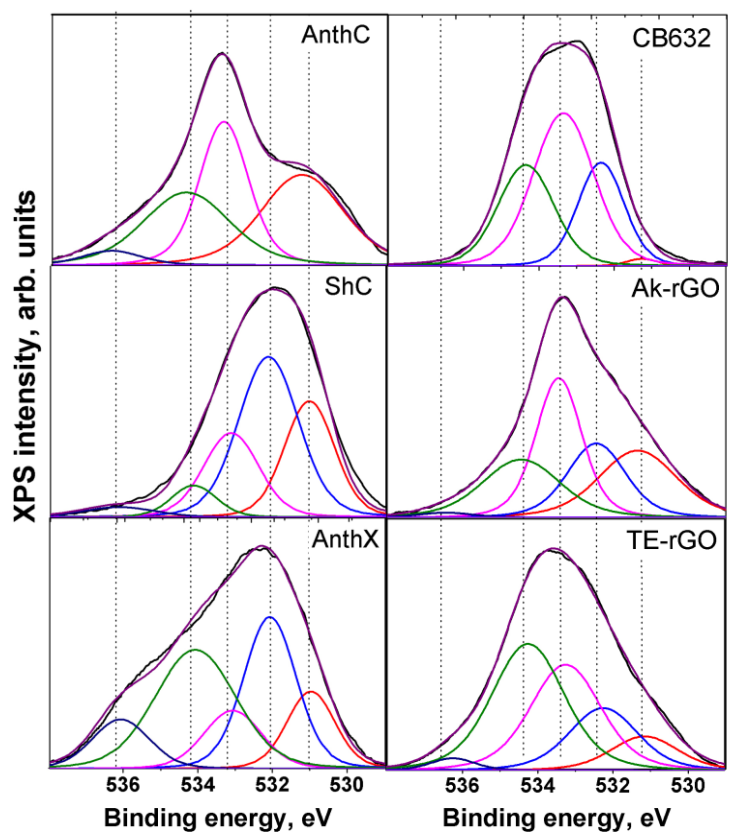

**Figure 7.** Expanded O 1s XPS spectra and fraction-distributions of Voigt-fitting-function peaks of O 1s spectra over peaks number of natural (left) and synthetic (right) amorphous carbons at room temperature, respectively. Spectra of graphite, AnthX, ShC, CB624, and CB632 were obtained previously [37] while those of AnthC, Ak-rGO, and TE-rGO were registered in the current study.

given elsewhere [37]. Hereinafter we shall restrict ourselves with O 1s spectra only that are the most informative for the case. Figure 7 presents a collection of pristine O 1s spectra obtained (black curves) alongside with resulted envelopes (dotted curves) of series of Voigt-fitting-function (VFF) peaks, which play the role of GFs in optical vibrational spectra and can be named as ***group binding energies*** (GBEs) analogously, presenting the asymptotic expansion of the pristine

**Table 4.** List of group binding energies of O 1s spectra attributed to amorphous carbons (composed from the published data [13])

| GBEs | BE, eV | Assignments |
|---|---|---|
| 1 | 531.2 | C=**O**, **O**=C-O-C=**O**, **O**=C-O-C (lactones and pairs of lactones) |
| 2 | 532.2 | O=C-**O**-C (lactones); **O**=C-C=**O** (o-quinones); **O**=C-OH; C=**O** in aggregated cyclic ethers with lactone |
| 3 | 533.2 | $sp^2$C-**O**H; C-**O**-C in cyclic ethers; C-O-C-**O**H (hydroxypyrans: singles and pairs ); O=C-**O**-C (lactones and pairs of lactones); O=C-**O**H; C-**O**-C in aggregated cyclic ethers with lactones |
| 4 | 534.2 | C-**O**-C in aggregated cyclic ethers; C-**O**-C-OH (hydroxypyran: singles and pairs); C-O-C in aggregated cyclic ethers with lactones |
| 5 | 536.2 | O=C-O-C-**O**-C-**O**-C-O-C=O in aggregated cyclic ethers with lactones |

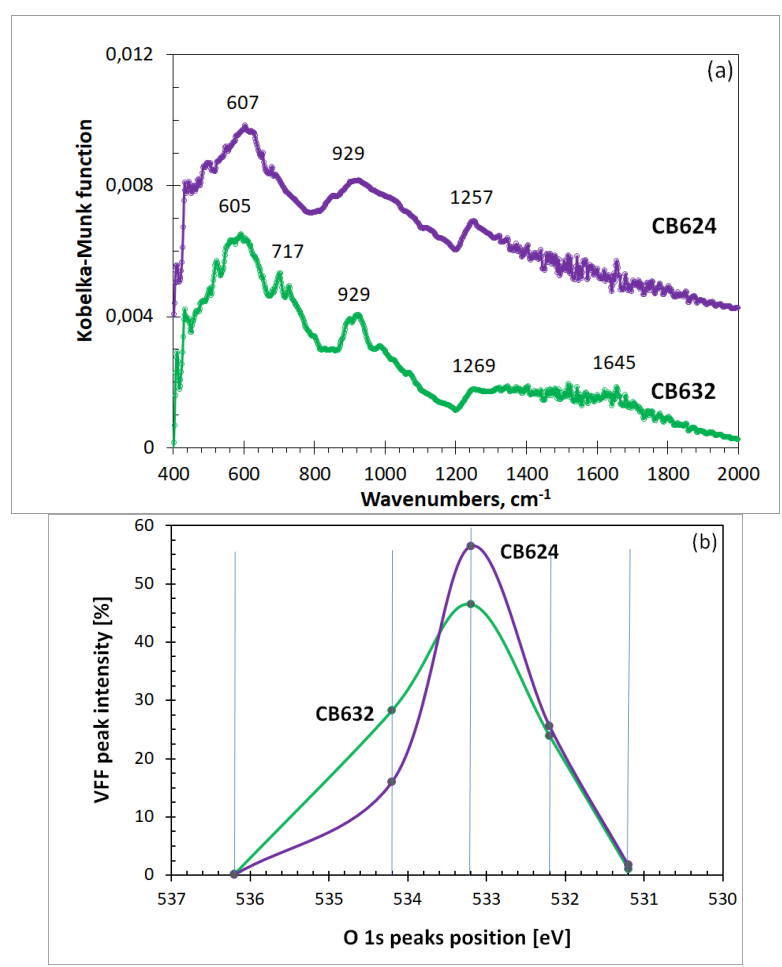

**Figure 8.** A combined DRIFT(a)-XPS(b) analysis of vibrational spectrum of technical graphenes.

spectra. The spectra are divided into two groups related to natural and synthetic ACs. To simplify further comparison, the pristine spectra of both groups are normalized per peak maxima. As seen in the figure, O 1s spectra clearly show that the oxygen chemistry in the framing areas of the studied ACs BSUs is individual for each species and quite markedly varies over the community. The composition of oxygen-containing groups (OCGs) was established by detailed analysis of the sample XPS spectra in the framework of the multi-complex-OCG (MC-OCG) approach successfully applied to the analysis of the composition of heat treated GO [13]. The collection of characteristic

GBEs and their assignment are listed in Table 4. As shown earlier [37], an analysis of the OCGs content attributed to the studied samples is better to perform using fractional contributions (FC) of each of the VFF component (GBE) into the total intensity of the O 1s spectra. Hereinafter we come back to this analysis completing it by simultaneous consideration of DRIFT spectra.

Figure 8 opens the series by the presentation of DRIFT-XPS picture of the studied carbon blacks. As seen in Fig. 8b, O 1s spectra of both samples are composed of GBEs 2, 3, and 4, among which GBE 3 evidently dominates. A comparison of the two spectra indicates that, generally, the composition of both sample is quite similar and the largest difference concerns the GBE 4 contribution. According to Table 4, the assignments offers a wide range of different OCGs to choose from hydroxyls; cyclic ethers; hydroxypyrans, lactones and pairs of lactones as well as aggregated cyclic ethers with lactones. Let us look for these groups in the DRIFT spectra of the species. As seen in Fig. 8a, at first glance, the presented DRIFT spectra look quite similar, indeed. Actually, three sharp features at ~600, 950 and 1260 $cm^{-1}$ of similar relative intensities form the spectra image. The feature is quite consistent with the fact that GBE 3 dominates in the O 1s spectra of both samples pointing to their common composition. The next consistency between DRIFT and XPS spectra concerns the absence of C=O groups. As seen in Fig. 8b, the contribution of GBE 1 in both spectra is practically nil, which evidences the absence of such groups in the studied samples. DRIFT spectra fully support the issue. According to Table 3, the carbonyl (C, O) GFs related to BSU molecules under study should be seen in the region above 1600 $cm^{-1}$. A weak hint of the corresponding feature is visible in the CB632 spectrum while the CB624 spectrum is practically blank in the region. Actually, according to both XPS and DRIFT spectra, C=O bonds are not characteristic for the structure of these ACs. On the other hand, small contributions of GBE 5 in O 1s spectra points to the minority of aggregated cyclic ethers. As seen in Table 4, exclusion of carbonyls and minority of aggregated cyclic ethers allows concluding that GBE 2 witnesses the presence of lactones while GBE 4 does the same concerning mainly hydroxyl pyran. The coincidence of GBE 2 contributions to spectra of both samples means that there is a comparable amount of lactones in their composition. On the other hand, a big difference of GBE 4 contributions evidences a significant preference for hydroxypyran in CB632 structure. This conclusion is consistent with higher hydrogenation of the sample with respect to CB624. The final conclusion about the chemical composition of the framing areas of the two samples can be made analyzing suggested assignments related to GBE 3. Since the characteristic is common for both CB624 and CB632 and due to practically nil hydrogenation of the former, confirmed by INS, $sp^2$C-**O**H and O=C-**O**H groups should be excluded from the consideration pointing that neither single hydroxyls nor single carboxyls take part in the BSU framings of both samples. Accordingly, cyclic ethers (predominantly), aggregated cyclic ethers and aggregated cyclic ethers with lactones lay the foundation of the CB624 BSU structure while hydroxypyran completes this composition in the case of CB632. Therefore, applying to DRIFT spectra of the samples, we have to analyze them from the viewpoint of C-O-C groups of cyclic ethers *et.al.* and (C-O-C)-OH groups of hydroxypyrans.

As well known, the triplet of bands is highly characteristic for cyclic esters, for which bands at 1260 and 970 $cm^{-1}$ present a highly characteristic doublet of the C-O stretchings while that one at 605 $cm^{-1}$ presents C-O deformations [51]. It is these three bands, which represent the spectrum of CB624. As for spectrum of CB632, there is only a weak hint of the 1260 $cm^{-1}$ band, a broad band at 970 $cm^{-1}$ is significantly changed, which can be caused by simultaneous appearance of 960 $cm^{-1}$ ν C-O(H) and 960 $cm^{-1}$ ν C-O bands. Apparently, the same concerns the low-frequency band transforming it into a doublet. As said earlier, the presence of hydroxyls in the framing area of CB632 BSU must provide not nil INS intensity. The INS spectrum of CB632 in Fig. 4 has not been interpreted until now. However, the suggestion about the presence of hydroxypyran in the framing area of the sample BSUs makes to draw attention to a weak low-frequency band below

100 cm$^{-1}$, which is characteristic for OH-containing species revealed by INS registered at NERA spectrometer [73].

After performed analysis of the hydrogen-pure carbon blacks, it seem to be reasonable to apply DRIFT-XPS approach to the analysis of the oxygen component of natural ACs and TGs. A comparative DRIFT-XPS picture consideration of natural amorphics is presented in Figs. 9a and 9b. As seen in Fig. 9b, the species, with so similar hydrogen components are absolutely different from the oxygen-content viewpoint. The only common feature joining them concerns carbonyl C=O groups, the presence of which in the BSU framings is justified by GBE 1 and (C, O) GF at ~1600 cm$^{-1}$. As for other contributions, which are lactones in ShC, lactones and aggregated cyclic ethers with lactones in AnthX, and cyclic ethers and carboxyls in AnthC, the characteristic GF structure typical for CB624 was not distinguished in DRIFT spectra of the samples. Apparently, the oxygen component cannot be detected in the DRIFT spectra of natural amorphics because its intensity constitutes not more than 10% due to which the corresponding bands get lost in the noise accompanying the intense DRIFT spectra and only C=O bands can be seen against the low background. However, the presence of carboxyls in the BSU framing of AnthC is strongly supported by the presence of the OH-characteristic low-frequency band in the INS spectrum in Fig. 4.

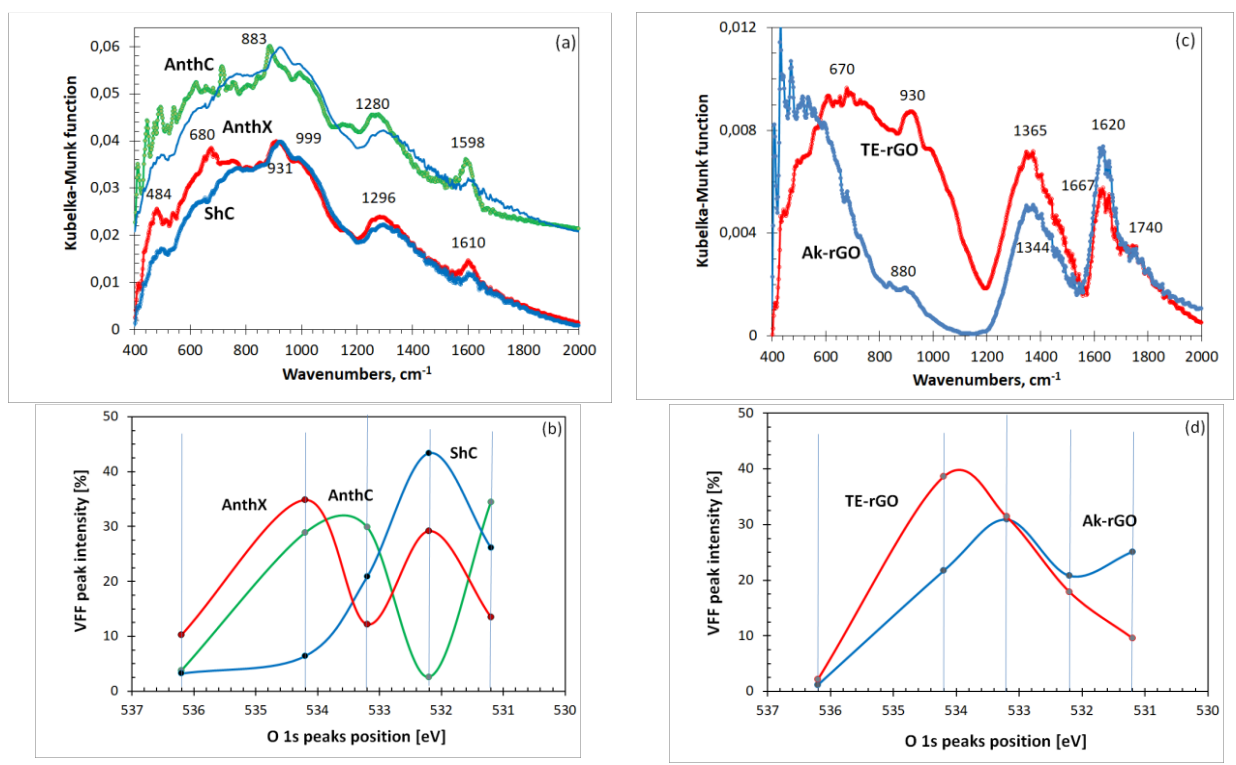

**Figure 9.** A combined DRIFT(a, c)-XPS(b,d) analysis of vibrational spectrum of natural amorphous carbons (left) and carbon blacks (right).

As for TGs, the oxygen component in their DRIFT spectra in Fig. 9c is more evident. Certainly, it is connected with a considerable lowering of the spectra intensity with respect to natural amorphics. As followed from the FC image of XPS spectra in Fig. 9d, DRIFT spectra of both samples should involve characteristic bands related to carbonyl groups, which are clearly seen in the region of 1600-1700 cm$^{-1}$. Besides, the benzyl-radical termination in the BSU circumference of TE-rGO is highly radicalized, which under ambient conditions is usually inhibited by hydroxyl addition, thus transforming benzyl radical into benzyl radical alcohol [59, 74]. We suppose that increasing of GBE 4 in the XPS spectrum of TE-rGO with respect to Ak-rGO convincingly evidences this fact. The issue is confirmed by a clearly seen feature at ~100 cm$^{-1}$ in the INS spectrum of the sample. According to the data of Table 3, one has to expect a considerable contribution in the DRIFT spectrum of the species at 960 and 1280 cm$^{-1}$. As seen in Fig. 9c, the expectations are fully met by the appearance of a band at 930 cm$^{-1}$ and a remarkable contribution to the band at 1365 cm$^{-1}$. Unfortunately, possibly present structural groups of lactones and cyclic ethers cannot be detected by DRIFT.

Summing up the comparative analysis of DRIFT spectra of the studied ACs, we can conclude the following. Predominant for natural ACs and comparable for TGs, the hydrogen component contributes to the DRIFT spectra shape and is well described in terms of GFs associated with *sp$^2$*C-H, *sp$^2$*C-CH$_2$, and *sp$^2$*C-CH$_3$ GFs. The structure of this component has a clearly visible specific molecular character, due to which the spectra of natural ACs, TE-rGO and Ak-rGO TGs have a benzene-like, benzyl-radical-like, and toluene-like character, respectively. In contrast, the oxygen component governs DRIFT spectra of carbon blacks while constitutes about 50% and 10% of the spectra of technical graphenes and natural amorphics, respectively. Accordingly, GFs related to different *sp$^2$*C-O as well *sp$^2$*C=O GFs lay in the foundation of the spectra description. A detailed comparable investigation of the *sp$^2$*C-O bond character of DRIFT and XPS spectra of ACs allowed distinguishing the presence of hydroxypyran and carboxyl structure compositions in the BSU framing areas of carbon black CB632 and anthracite, respectively, which made it possible to complement the hydrogen components of the species with hydroxyl torsions around *sp$^2$*C-OH bonds. The presence of the relevant vibrations was supported by the INS study.

## 7. Models of BSU molecules of amorphous carbons

The obtained description of DRIFT spectra in terms of GFs allows addressing atomic structures of the BSU framing areas of the studied ACs. The first attempt to construct the relevant molecular models was made earlier [37]. The results of the current study makes it possible to support and/or correct the previous models and to suggest previously absent molecular structures of anthracite and technical graphenes. As previously, a naked (5, 5)NGr molecule of ~1.5 nm in size is chosen as a template for models related to natural amorphics an carbon black CB632. A bigger (9, 9)NGr molecules (see inset in Fig. 6) is taken as template for molecular models of technical graphenes and black carbon CB624. The main goal of the model presentation is not to suggest exact replica of BSUs of the studied ACs, the evident extreme complexity and variety of which does not so far allow getting exact solution. The aim is to give a presentation of which atomic compositions can be met on this way under restricted conditions concerning size and general chemical content of the molecules accumulated in Tables 1 and 2.

Begin with the description of the chemical content related to shungite carbon, presented in Table 8 of the previous study [37]]:

**ShC**: carbonyls sp$^2$C=O; *acid anhydride O=C-O-C=O*; *o*-quinone O=sp$^2$C-sp$^2$C=O, *carboxyls sp$^2$C=OOH*. Methine groups.

Recall that the description resulted from the analysis of XPS spectrum. According to the data obtained in the current study, nothing new should be added to the list while, in contrast, acid anhydride and carboxyls should be excluded. The first – due to the absence of the corresponding GF in the DRIFT spectrum at ~1800 cm$^{-1}$, and the second – as not supported with INS data. Thus, preserving "C=O" carbonyl character of the amorphic, the models, presented in Fig. 10a, can give a clear vision of chemical compositions that comply with the data in Table 2. Methine groups are added as well to take into account the hydrogen component. Six hydrogen and four oxygen atoms implement the chemical content of ShC BSU in Table 2 with respect to 66 carbon atoms of the template molecule (5, 5)NGr. The composition presented in Fig. 10a evidently is not the only one and a set of carbonyl-ketone configurations is possible, subordinate only to the hydrogen/oxygen ($\mathcal{HO}$) number "$6\mathcal{H}$ and $4\mathcal{O}$". It should be noted that the first model of the ShC BSU, generally fully identical to presented in the figure, was proposed long before the results of analytical studies, basing on a quantum-chemical analysis of the redox processes underlying the formation of this mineral [20, 42, 75]. Close identity of the models can be considered as a serious argument in favor of the idea of the nucleation and accumulation of ShC BSUs in due course of redox reactions with the participation of graphene lamellas in a hydrothermal medium, firstly proposed in [42].

As for anthraxolite, presented earlier [37] says:

**AnthX:** hydroxyls sp$^2$C-OH; C-O-C-OH (hydroxy pyran-HP) and pairs of HPs; C=OOC(lactone) and pairs of lactones; aggregated cyclic ether with lactone. Methine groups,

The current study results confidently permit to exclude hydroxyls and hydroxyl pyran from the list, leaving "O=C-O-C" lactone character of the species. The HO number constitutes "10H and 4 O". Accordingly, the AnthX model presented in Fig. 10b involves 10 methine groups and a composition of aggregated cyclic ethers with lactone. The methine group combination is fixed while it is easy to imagine other possible combinations of lactone (s) and cyclic ethers formed by 4 oxygen atoms.

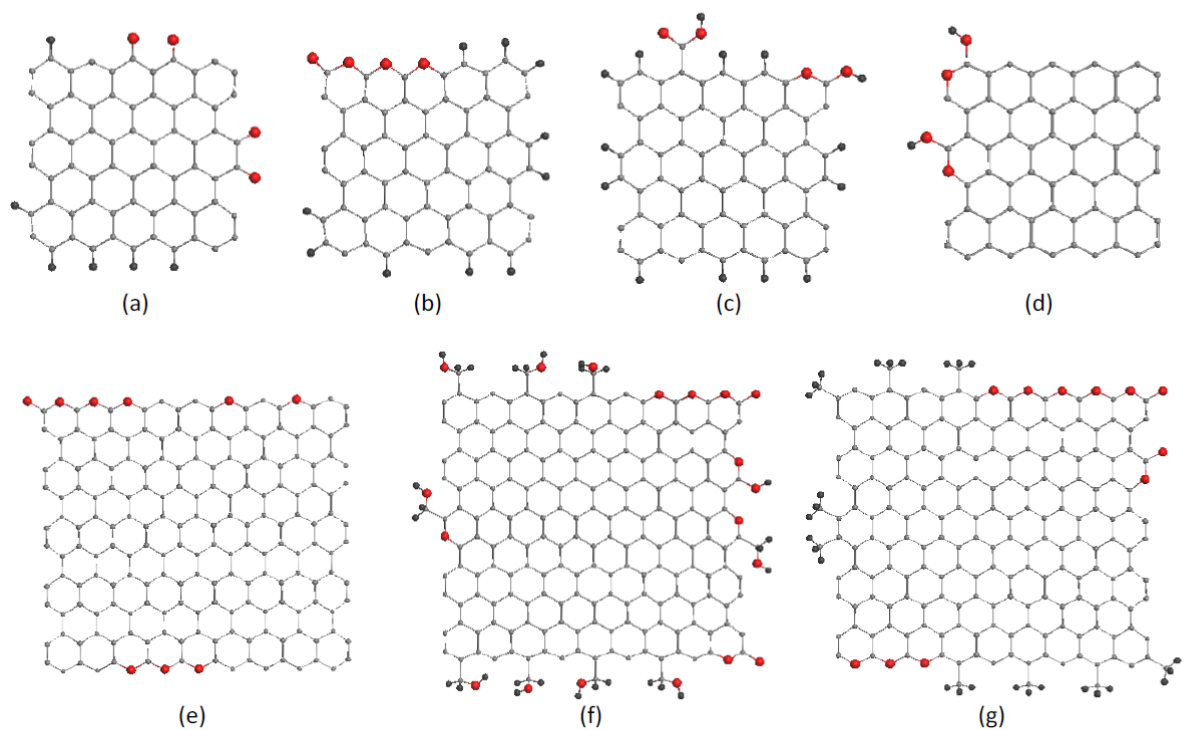

**Figure 10.** Molecular models of basic structure units of amorphous carbons: shungite carbon (a), anthraxolite (b), anthracite (c), carbon black CB632 (d), carbon black CB624 (e), and technical graphenes TE-rGO (f) and Ak-rGO (g) (see text).

Anthracite is a new species in the natural amorphics list and its presentation in light of current study looks the following:

**AnthC:** carboxyls sp$^2$C-COOH; cyclic ethers; aggregated cyclic ethers; pyran and hydroxypyran. Methine groups

By analogy with ShC and AnthX, anthracite character can be marked as "C-(O:C)-OH" where (O:C) means either single bond of pyran or double bond of carboxyl. The $\mathcal{HO}$ number for (5, 5)NGr molecule is "14$\mathcal{H}$ and 4$O$". One of the possible model, involving 14 methine groups, one carboxyl and one hydroxypyran, is presented in Fig. 10c. Other models involving four oxygen atom combinations of hydroxyl (s), pyran (s), and cyclic ethers are possible as well.

Carbon black CB632 completes the consideration of ACs with small size of the relevant BSUs. Previously concluded [37] its presentation stays the following:

**CB632**: C-O-C in cyclic ether; aggregated cyclic ethers; C-O-C of pairs of cyclic ether and lactone and/or hydroxypyran.

The current study leaves the first part of the conclusion unchanged while makes a particular emphasis on hydroxypyran in the last part. The model shown in Fig. 10d presents the strongest composition of this type consisting of two hydroxypyrans. In practice, this composition evidently does not dominate and gives way to four-oxygen-atoms single cyclic ethers, as well as cyclic ethers with lactone.

Concluding the description of small-size BSUs, is necessary to note that small size of the molecule as well as small number of oxygen atoms in their circumference do not allow, within the framework of one model, to reveal all possible combinations of these atoms, which are manifested in the XPS spectra. This means that, in reality, BSUs of all natural amorphics and CB632 are not standard but are sets of varied structures characterized by the average $\mathcal{HO}$ number.

Carbon black CB624 open the presentation of big-size BSUs. As previously showed [37], variations of its chemical compositions are quite limited.

**CB624**: C-O-C in cyclic ester, aggregated cyclic ester and aggregated cyclic ether with lactone.

The current study fully supports this conclusion. The $\mathcal{HO}$ number is (0$\mathcal{H}$ and 9$O$) and represents, as previously, the atomic contribution based on the data in Table 2 with respect to 190-atom (9, 9)NGr template molecule. A bigger platform of the molecule allows presenting listed configurations on the example of one model shown in Fig. 10e.

Technical graphenes were not modelled before and model structures, presented in Fig. 10, are suggested based on the current study results. Analogously presentations given above, TE-rGO BSU models should involve the following compositions:

**TE-rGO**: aggregated cyclic ethers and aggregated cyclic ethers with lactones; hydroxypyrans and lactones, both singles and pairs. Hydroxymethyl groups.

The characteristic $\mathcal{HO}$ number constitute (30$\mathcal{H}$ and 20$O$). The model in Fig. 10f involves all the types of the above compositions. Sure, the compositions themselves can vary keeping $\mathcal{HO}$ number fixed.

The chemical compositions of the Ak-rGO BSUs are rich as well but differ from those of TE-rGO ones by the absence of hydroxyl containing groups. Accordingly, the description of the Ak-rGO BSUs is the following:

**Ak-rGO**: aggregated cyclic ethers and aggregated cyclic ethers with lactones; lactones and pairs of lactones. Methyl groups.

As in the previous case, big size of the template molecule allows to present all the types of possible oxygen-containing compositions, subordinate to the $\mathcal{HO}$ number (27$\mathcal{H}$ and 11$O$), on one model in Fig. 10g.

As said earlier, models shown in Fig. 10 do not pretend on presentation of exact BSU molecules since they give averaged picture strictly subordinate to BSU size and chemical content of the solid materials. A lot of new information is still necessary to give a preference to one or another functional group. It will take a lot of time and painstaking research to find the sources of the required new information.

## 8. Conclusion

The paper presents a comparative analysis of INS, DRIFT and XPS spectra of a set of amorphous carbons consisting of three representatives of natural substances (shungite carbon, anthraxolite and anthracite), two technical graphenes and two industrial products of carbon black. A combined application of INS and DRIFT study allowed exhibiting a hydrogen component of the ACs vibrational spectra. A combined consideration of XPS and DRIFT spectra allowed the detection of the corresponding oxygen component. This analysis was aimed at establishing the atomic structure of the circumference of graphene molecules, which are the basic structure units of the studied amorphics. It was found that this area is compositionally different for different samples and that this difference is determined by the chemical history of the products. So, in industrial samples, hydrogen is almost completely absent and the framing of their BSUs is carried out by oxygen-containing groups. On the contrary, methine groups represent the hydrogen component of natural amorphics in all cases. A particular "riding effect", characteristic for these groups in graphene molecules and responsible for the enhancement of electrooptic characteristics of carbon atoms vibrations, has been discovered. The framing area of technical graphenes contains a comparable participation of both components, but the hydrogen component is represented by either hydroxymethyl or methyl groups, depending on the methods of the product preparation. The results obtained collectively show that amorphous carbon is not a standardized material and is individual in nature, depending on the history of its origin or production. Because of this, each application of this material must be accompanied by a thorough analysis of individual properties. The only common property of all amorphics is the radical nature of their molecular structural elements [37], the disclosure or use of which depends on other individual characteristics.

**Acknowledgements**

The authors are thankful to V.A. Benderskiy and N.N.Rozhkova for fruitful discussions, to S.V.Tkachev and V.M.Mel'nikov for supplying with samples of technical graphenes and necessary information concerning the sample production. The work has been performed using the equipment of the Multi-User Analytical Center (Institute of Problems of Chemical Physics, RAS) and Scientific Center in Chernogolovka RAS in accordance with the state task, state registration No. AAAA-A19-119061890019-5. The study was partially carried out using the equipment of the "Khimiya" Common Use Centre (Institute of Chemistry, Komi Science Centre, Ural Branch, RAS). The publication has been prepared with the support of the "RUDN University Program 5-100".

**Statement about conflicts**

There are no conflicts between the authors